\documentclass[pra,twoside,twocolumn,aps,amsmath,amssymb,amsfonts]{revtex4-1}

\usepackage{amsmath,bbm,mathrsfs}
\usepackage{amsthm}
\usepackage{graphicx}
\usepackage{amsfonts}
\usepackage{amssymb}
\usepackage{color}
\usepackage{polski}
\usepackage{enumitem}
\definecolor{myurlcolor}{rgb}{0,0,0.7}
\definecolor{myrefcolor}{rgb}{0.8,0,0}
\usepackage{hyperref}
\hypersetup{colorlinks,
linkcolor=myrefcolor,
citecolor=myurlcolor,
urlcolor=myurlcolor}

\def\textbf#1{{\bf #1}}
\def\be{\begin{equation}}
\def\ee{\end{equation}}
\def\ben{\begin{eqnarray}}
\def\een{\end{eqnarray}}
\def\eea{\end{array}}
\def\bea{\begin{array}}
\newcommand{\Tr}[0]{\mathrm{Tr}}
\newcommand{\ot}[0]{\otimes}
\newcommand{\bei}{\begin{itemize}}
\newcommand{\eei}{\end{itemize}}
\newcommand{\ket}[1]{|#1\rangle}
\newcommand{\bra}[1]{\langle#1|}

\newcommand{\proj}[1]{\ket{#1}\!\bra{#1}}

\newcommand{\spa}[0]{\mathrm{span}}

\newtheorem{thm}{Theorem}

\theoremstyle{definition}

\begin{document}
%

\title{Optimal decomposable witnesses without the spanning property}

\author{Remigiusz Augusiak$^{1}$, Gniewomir Sarbicki$^{2}$, Maciej Lewenstein$^{1,3}$}

\affiliation{$^{1}$ICFO--Institut de Ci\`encies Fot\`oniques,
E--08860 Castelldefels (Barcelona), Spain}

\affiliation{$^{2}$Insitute of Physics, Nicolaus Copernicus
University, Grudzi
\k{a}dzka 5, 87-100 Toru\'n, Poland}

\affiliation{$^{3}$ICREA--Instituci\'o Catalana de Recerca i
Estudis Avan\c{c}ats, Lluis Companys 23, 08010 Barcelona, Spain}

\begin{abstract}
One of the unsolved problems in the characterization of the
optimal entanglement witnesses is the existence of optimal
witnesses acting on bipartite Hilbert spaces
$\mathcal{H}_{m,n}=\mathbbm{C}^{m}\ot\mathbbm{C}^{n}$ such that
the product vectors obeying $\langle e,f|W|e,f\rangle=0$ {\it do
not} span $\mathcal{H}_{m,n}$. So far, the only known examples of such witnesses were found among indecomposable witnesses, one of them being the witness corresponding to the Choi map. However, it remains
an open question whether decomposable witnesses exist without the property of spanning. Here we answer this question
affirmatively, providing systematic examples of such witnesses. Then, we generalize some of the recently obtained results on the characterization of $2\ot n$ optimal decomposable witnesses [R. Augusiak {\it et al.}, \href{http://iopscience.iop.org/1751-8121/44/21/212001}{J. Phys. A {\bf 44}, 212001 (2011)}] to finite-dimensional Hilbert spaces $\mathcal{H}_{m,n}$ with $m,n\geq 3$.
\end{abstract}

\maketitle

\section{Introduction}

Characterization and classification of entanglement witnesses
(EWs) remains an unsolved problem in entanglement theory
\cite{HorodeckiRMP,GuhneToth}. Recall that by an entanglement
witness \cite{Horodecki96PLA}, we understand a Hermitian operator,
usually denoted by $W$, acting on some bipartite Hilbert space
$\mathbbm{C}^{m}\ot \mathbbm{C}^{n}$ such that $\langle
W\rangle_{\sigma}\geq 0$ for any separable state \cite{WernerSep}
$\sigma$ acting on $\mathbbm{C}^{m}\ot \mathbbm{C}^{n}$, while for
some entangled $\varrho$, $\langle W\rangle_{\varrho}<0$. In the
latter case, we say briefly that entanglement of $\varrho$ or just
$\varrho$ is detected by the EW $W$, hence the term
\cite{Terhal00PLA}. What makes these objects important from
an entanglement detection point of view is that (as proven in Refs.
\cite{Horodecki96PLA} and \cite{Horodecki01PLA}) for any entangled state (bipartite or multipartite) there exists some EW detecting it
(i.e., having negative mean value in this state). Much effort has
been put toward designing EWs that detect entanglement in various
bipartite and multipartite physical systems (see, e.g., Ref.
\cite{constr}). They also allow for the quantitative analysis of entanglement (see, e.g., Ref. \cite{ograniczenia}), and, finally, since these objects are just quantum observables, the qualitative and quantitative detection of entanglement is feasible from the experimental point of view \cite{exp}.

Particularly important for the characterization of EWs is the
notion of optimality introduced in Ref. \cite{optimization} (see
also Ref. \cite{Terhal2} for a state-dependent definition
of optimality). Roughly speaking, optimal EWs are the ones that
detect (in the set-theoretic terms) the largest set of entangled
states. In other words, a given EW $W$ is optimal if there
exist no other witness detecting the same states as $W$ and
additionally some states which are not detected by $W$. As every
witness can be optimized \cite{optimization}, it happens that the
optimal EWs are sufficient to detect all the entangled states. It
is then of significant importance to isolate and characterize the
set of optimal EWs.

Many interesting results have been obtained in research
aiming to realize this goal (see, e.g., Refs. \cite{prop} and \cite{Korbicz}). However, though this effort, the characterization of optimal EWs is far from being accomplished, and our knowledge about their structure remains unsatisfactory.

Very recently, in Ref. \cite{nasza}, some of us have have provided
a more exhaustive characterization of all qubit-qunit decomposable
EWs (DEWs). It was shown that product vectors orthogonal to any
completely entangled subspace (CES) of
$\mathbbm{C}^{2}\ot\mathbbm{C}^{n}$, after being partially
conjugated, span the latter. On the level of DEWs, together with
results already established in Ref. \cite{optimization}, this means that for any DEW $W$ acting on $\mathbbm{C}^{2}\ot\mathbbm{C}^{n}$ the
following two equivalences hold: (i) $W$ is optimal if and only if (iff) it takes the form $W=Q^{\Gamma}$ with $Q\geq 0$ supported on some CES and (ii) $W$ is optimal iff the qubit-qunit product vectors satisfying $\langle e,f|W|e,f\rangle=0$ span $\mathbbm{C}^{2}\ot\mathbbm{C}^{n}$. Since it is important for our considerations, let us refer to the latter property as {\it the property of spanning}.

On the other hand, it was shown in Ref. \cite{nasza} that already
the two-qutrit DEWs do not follow the above characterization. More
precisely, while for all DEWs $W=Q^{\Gamma}$ with $Q$ of rank 1
or 2, the above equivalences also hold, in the case when
$r(Q)=3,4$ either there exist non-optimal DEWs taking the above
form or there exist optimal witnesses without the property of
spanning. While the former question has very recently been
solved \cite{Kye}, the latter one seems particularly interesting
and has not been answered so far. The existence of optimal EWs
without the property of spanning is already known in the case of
indecomposable witnesses, where one has the so-called Choi
witness, that is, an indecomposable EW acting on
$\mathbbm{C}^3\ot\mathbbm{C}^3$ generated from the known Choi map
\cite{Choi,ChoiLam}. (For the proof that this witness does not have
the property of spanning, see Ref. \cite{Korbicz}.) This fact,
however, remains unknown in the decomposable case.

The primary purpose of this paper is to clarify this point and
show that for any Hilbert space
$\mathbbm{C}^{m}\ot\mathbbm{C}^{n}$, obviously except for
$\mathbbm{C}^{2}\ot\mathbbm{C}^{n}$, there always exist such
witnesses. The secondary aim is to discuss possible
generalizations of the results obtained in Ref. \cite{nasza} to
higher-dimensional DEWs. In particular, we show that under some
condition all CESs supported in $\mathbbm{C}^n\ot\mathbbm{C}^n$
of dimension $n-1$ have the property of spanning. Then, we prove that any CES in $\mathcal{H}_{n}$ of dimension less than
$n-1$ can be extended (i.e., is a subspace) to some
$(n-1)$-dimensional CES and therefore also inherits this
property.

We apply this analysis to the CESs supported in
$\mathbbm{C}^{4}\ot\mathbbm{C}^{4}$. We show that all such CESs of
dimension 3 obey the above condition. Consequently, for all
two-ququart DEWs $W=Q^{\Gamma}$ with $r(Q)\leq 3$, the above
equivalences also hold. Together with the already obtained
results for two-qubit and two-qutrit DEWs \cite{nasza}, this suggests to conjecture that the above equivalences are valid for DEWs
$W=Q^{\Gamma}$ acting on $\mathbbm{C}^n\ot\mathbbm{C}^n$ with
$r(Q)\leq n-1$. However, we provide examples of
$(n-1)$-dimensional CESs supported in
$\mathbbm{C}^m\ot\mathbbm{C}^n$ with $3\leq m\leq n$ without the
property of spanning, meaning that the local dimensions play some
role in the above conjecture.

The paper is organized as follows. In Sec. \ref{Preliminaries}, we
recall basic notions regarding witnesses and optimality. In Sec.
\ref{Examples}, we present our main results, that is, examples of
optimal DEWs without the property of spanning. In Sec. \ref{Further}, we discuss
the general situation with respect to characterization of optimal
DEWs. We conclude with Sec. \ref{Conclusion} and outline
possible directions for further research.

\section{Preliminaries and the problem}
\label{Preliminaries}

First, we set the notation and recall the basic
notions and known facts regarding EWs and, in particular,    DEWs.

By $\mathcal{H}_{m,n}$, $\mathbbm{M}_{m,n}$, and $\mathbbm{D}_{m,n}$
we denote, respectively, the product Hilbert space
$\mathbbm{C}^{m}\ot\mathbbm{C}^{n}$, the set
$\mathbbm{M}_{m}(\mathbbm{C})\ot\mathbbm{M}_{n}(\mathbbm{C})$ with
$\mathbbm{M}_d(\mathbbm{C})$ standing for the set of $d\times d$
complex matrices, and finally the subset of positive elements of
$\mathbbm{M}_{m,n}$ with unit trace. In the case when the
dimensions of both subsystems are equal, we use shorter
notation: $\mathcal{H}_{m}$, $\mathbbm{M}_{m}$, and
$\mathbbm{D}_m$, respectively. By the Schmidt rank of a pure state
$\ket{\psi}\in\mathcal{H}_{m,n}$ we understand the rank of a
density matrix of one of the subsystems of $\ket{\psi}$. Then, we
say that a subspace $V$ of $\mathbbm{C}^{m}\ot\mathbbm{C}^n$ is
supported in the latter if $V$ cannot be embedded in some Hilbert
space $V_1\ot V_2$, where either $V_1$ or $V_2$ is a proper
subspace of $\mathbbm{C}^{m}$ or $\mathbbm{C}^{n}$, respectively.
Finally, we often denote vectors from $\mathbbm{C}^m$ in
the following way:
\begin{equation}
\mathbbm{C}^{m}\ni
\ket{f}=\sum_{i=0}^{m-1}f_{i}\ket{i}=(f_{0},f_{1},\ldots,f_{m-1}).
\end{equation}

By an {\it entanglement witness} (EW), we understand a Hermitian
operator $W\in \mathbbm{M}_{m,n}$ which is block-positive, i.e.,
such that $\langle e,f|W|e,f\rangle\geq 0$ holds for any product
vector $\ket{e,f}\in \mathcal{H}_{m,n}$, and there exist entangled
states $\rho$ for which $\Tr(W\rho)<0$.

We call an EW $W$ {\it decomposable} \cite{optimization} if it can
be written as
\begin{equation}\label{DEWs}
W=P+Q^{\Gamma},
\end{equation}
with $P$ and $Q$ being positive operators. EWs which do not admit
this form are called {\it indecomposable}. From now on we restrict
our attention to decomposable EWs since our results concern only
this subset.

We can now turn to the notion of optimality. For this purpose, for
a given EW $W\in\mathbbm{M}_{m,n}$ let us introduce the following
sets
\begin{equation}
  \mathcal{P}_W=\{\ket{e,f}\in\mathcal{H}_{m,n}|\langle e,f|W|e,f\rangle=0\}
\end{equation}
and
\begin{equation}
\mathcal{D}_W=\{\varrho \in\mathbbm{D}_{m,n}| \Tr(\varrho W)<0\}.
\end{equation}
Now, taking two EWs $W_i$ $(i=1,2)$, we say that $W_1$ is {\it
finer} that $W_2$ if $\mathcal{D}_{W_2}\subseteq
\mathcal{D}_{W_1}$. Then, if there does not exist any witness
which is finer than $W\in\mathbbm{M}_{m,n}$, we call it {\it
optimal}. That is optimal EWs are those that are maximal with
respect to the above relation of inclusion.

In Ref. \cite{optimization}, it was shown that an EW
$W\in\mathbbm{M}_{m,n}$ is optimal if and only if the matrix
$\widetilde{W}(\epsilon,P)=W-\epsilon P$ is no longer
block-positive for any $\epsilon>0$ and $P\geq 0$. Clearly, in
order to verify the optimality of a given EW $W$, it suffices to
check the above condition for positive $P$ supported on subspaces
that are orthogonal to $\mathcal{P}_W$ because for any
$P$ for which $PP_W\neq 0$ there always exists a product vector
$\ket{e,f}\in \mathcal{P}_W$ such that $\langle
e,f|P|e,f\rangle\neq 0$ and therefore $\langle e,f|W-\epsilon
P|e,f\rangle=-\langle e,f|P|e,f\rangle<0$. This fact
immediately implies a sufficient condition for optimality:
\begin{description}
\item[(i)] If $\mathcal{P}_{W}$ spans $\mathcal{H}_{m,n}$ then the EW $W\in\mathbbm{M}_{m,n}$ is optimal.
\end{description}
Since the latter property is directly related to the notion of
optimality, let us remember that we
refer to it as {\it property of spanning}.

Furthermore, application of this condition for optimality to
the DEWs (\ref{DEWs}) yields the necessary condition for
optimality of DEWs \cite{optimization}:
\begin{description}
\item[(ii)] If a given DEW $W\in\mathbbm{M}_{m,n}$ is optimal, then it has to be of the form
\begin{equation}\label{form}
W=Q^{\Gamma}, \qquad Q\geq 0 , \qquad Q\;\,
\mathrm{supported}\;\,\mathrm{on}\;\, \mathrm{CES},
\end{equation}
\end{description}
that is, $Q$ is a positive operator supported on some completely
entangled subspace.

It has been whether the opposite
implications of (i) and (ii) also hold. Clearly, a solution to
this problem is crucial from the point of view of a complete
characterization of optimal DEWs. As already said, both statements
become equivalences for all DEWs from $\mathbbm{M}_{2,n}$ and some
from $\mathbbm{M}_{3,3}$ \cite{nasza}. Quite surprisingly,
however, the existence of nonoptimal DEWs taking the form
(\ref{form}) has very recently been reported \cite{Kye} (see also
Ref. \cite{examples} in this context), proving that in general
the opposite statement to (ii) does not hold. In particular, it is
a consequence of the fact that generally one is able to decompose
$\mathcal{H}_{m,n}$ (except for $\mathcal{H}_{2,n}$ and
$\mathcal{H}_3$) as a direct sum of two CESs\footnote{Just to give
an example, consider the six-dimensional subspace of
$\mathcal{H}_{3,4}$ spanned by the vectors
$\{\ket{01}-\ket{10},\ket{02}-\ket{20},\ket{03}-(\ket{12}-\ket{21}),
\ket{13}-\ket{22},\ket{11}-\ket{23},\ket{00}-\ket{13}-\ket{21}\}$.
One checks that this subspace and its six-dimensional complement
do not contain any product vectors.} (see, e.g., Ref.
\cite{Leinaas}). Any such CES can support a DEW of the form
(\ref{form}), however, the latter cannot be tangent to the set of
separable states, and thus it is certainly not an optimal witness
\cite{Kye}.

The primary aim of the present paper is to disprove also
the inverse of (i) for general DEWs. For this purpose, we first
prove that some of the two-qutrit witnesses without the property
of spanning found in Ref. \cite{nasza} are optimal. Then, we
generalize these examples to every $\mathcal{H}_{m,n}$ with
$m,n\geq 3$, showing that every Hilbert space (except for
$\mathcal{H}_{2,n}$) admits optimal DEWs (in the sense that there
exist DEWs acting on this Hilbert space) without the property of
spanning.

It should also be emphasized that
our construction provides examples of optimal DEWs
without the property of spanning. Recall that in the
indecomposable case the only examples known so far arise
from the Choi map \cite{Choi}.

Another aim of our paper is to provide some generalizations of
the results of Ref. \cite{nasza} to higher-dimensional Hilbert
spaces $\mathcal{H}_{m,n}$.

Let us notice that we can look at the above problems from a
more general perspective. To determine the set $\mathcal{P}_W$ for
DEWs (\ref{form}), we need to find the product vectors
$\ket{e,f}\in\mathcal{H}_{m,n}$ obeying $0=\langle
e,f|Q^{\Gamma}|e,f\rangle=\langle e,f^{*}|Q|e,f^{*}\rangle$. Because $Q\geq 0$ and $\mathrm{supp}(Q)$ is a CES, the
problem of determining $\mathcal{P}_W$ reduces to the problem of
determining product vectors in $V^{\perp}$ for $V$ being a CES of
$\mathcal{H}_{m,n}$. Furthermore, it is fairly easy to see
that every CES $V$ admits (is the sense of being supported on $V$)
a positive $Q$ giving rise to a proper EW (that is
$Q^{\Gamma}\ngeqslant 0$). Consequently, instead of working in
terms of particular $Q$s we can work in terms of CESs of
$\mathcal{H}_{m,n}$. For any CES $V$ let us then define the analog
of $\mathcal{P}_W$ (defined for EWs):
\begin{equation}
\mathcal{P}_V=\{\ket{e,f^{*}}\in\mathcal{H}_{m,n}\,|\,\ket{e,f}\in
V^{\perp}\}.
\end{equation}

Recall that CESs were already investigated in the literature (see
Refs. \cite{Wallach,Parthasarathy,WalgateJPA08,CubittJMP08}), and
the largest dimension of a CES in
$\mathcal{H}_{m,n}$ is $(m-1)(n-1)$. This translates to an upper
bound on the possible ranks of $Q$ in Eq. (\ref{form}). In the
case of $\mathcal{H}_{2,n}$ this reduces to $n-1$ and all CESs of
dimension less or equal $n-1$ have the property of spanning
\cite{nasza}. For the higher-dimensional Hilbert spaces, our
result clearly establishes an upper bound on the largest dimension
for all CESs to have the property of spanning, which is $\dim V<
n$. We conjecture that this bound is tight when local dimensions
are equal, while we show that when $m<n$ there exist CESs of
dimension $n-1$ without the property of spanning.

\section{Completely entangled subspaces without the property of spanning}
\label{Examples}

Let us introduce some additional
notation. Clearly, any vector $\ket{x}\in\mathbbm{C}^{m}$ can be
written as $\ket{x}=x_0\ket{0}\oplus\ket{\widetilde{x}}$ with
$\ket{\widetilde{x}}\in \spa\{\ket{1},\ldots,\ket{m-1}\}$. In what
follows, $\ket{\widetilde{x}}$ always denotes a vector coming
from the above decomposition. Also, for any
$A:\mathbbm{C}^{m}\to\mathbbm{C}^{m}$, by $\widetilde{A}$ we
denote the $(m-1)\times (m-1)$ matrix obtained from $A$ by
removing its first row and first column.

\subsection{The case of equal dimensions}
\label{equal}

Let us start from the case of equal dimensions of both subsystems
and consider the $m$-dimensional subspace $V$ of $\mathcal{H}_m$
spanned by the vectors
\begin{equation}\label{PsiI}
\ket{\Psi_i}=a_i\ket{0}\ket{i}-b_i\ket{i}\ket{0}\qquad
(i=1,\ldots,m-1)
\end{equation}
with $a_i,b_i\neq 0$ and $\ket{\Psi_m}$ being so far any
non-product vector from
$(\mathrm{span}\{\ket{1},\ldots,\ket{m-1}\})^{\ot 2}$. It is useful to notice that via the vectors-matrices
isomorphism (see the Appendix), $\ket{\Psi_i}$ $(i=1,\ldots,m-1)$
correspond to matrices $A_i=a_i|0\rangle\!\langle
i|-b_i|i\rangle\!\langle 0|$ and $\ket{\Psi_m}$ correspond to some, so far
unspecified, matrix $A_m$ having the entries of the first row and
the first column all equal to zero and whose rank obeys $2\leq
r(A_m)\leq m-1$. For the sake of simplicity we assume that
$r(A_m)=m-1$ ($\ket{\Psi_m}$ has Schmidt rank $m-1$).

Before going into detail, it should be noticed that by a local
nonsingular transformation on one of the subsystems, the subspace
$V$ can be transformed to a subspace spanned by
$\ket{\Psi'_i}=\ket{0}\ket{i}-\ket{i}\ket{0}$ $(i=1,\ldots,m-1)$
and $\ket{\Psi'_m}$ which is again a nonproduct vector from
$(\mathrm{span}\{\ket{1},\ldots,\ket{m-1}\})^{\ot 2}$. Since local
nonsingular transformations do not influence our analysis at all,
in further considerations we can assume that $a_i=b_i=1$.

Finally, it is fairly easy to see that $V$ is a completely
entangled subspace (see the Appendix for the detailed proof).

Now, let us find all the product vectors $\ket{x,y}$
($\ket{x},\ket{y}\in\mathbbm{C}^{m}$) orthogonal to $V$. This can
be done by solving a system of $m$ linear homogenous equations
\begin{equation}\label{system}
\bra{\Psi_i}x,y\rangle=0 \qquad (i=1,\ldots,m),
\end{equation}
which can be restated as the system of equations for the vector
$\ket{y}$ of the form
\begin{equation}\label{equation}
M(x)\ket{y}=0
\end{equation}
with the $m\times m$ $\ket{x}$-dependent matrix $M(x)$ given by
\begin{equation}\label{macierzM}
M(x)=\left(
\begin{array}{ccccc}
x_1      & -x_0                        & 0      & \ldots & 0\\
x_2      & 0                          & -x_0    & \ldots & \vdots\\
\vdots   & \vdots                     & \vdots & \ddots & 0\\
x_{m-1}  & 0                          & \ldots & \ldots & -x_0\\
0        & \langle x^{*}|A^{*}_m|1\rangle & \ldots & \ldots &
\langle x^{*}|A^{*}_m|m-1\rangle
\end{array}
\right).
\end{equation}
Clearly, Eq. (\ref{equation}) has a nontrivial solution
iff $\det M(x)=0$. After simple algebra the latter can be
rewritten as
\begin{eqnarray}
\det M(x)&=&(-1)^{m}x_0^{m-1}\sum_{i,j=1}^{m-1}x_i
[A_m^{*}]_{ij}x_j\nonumber\\ &=&(-1)^{m}x_0^{m-1}\langle
x^{*}|A_m^{*}|x\rangle\nonumber\\
&=&0,
\end{eqnarray}
and is satisfied iff either $x_0=0$ or $\langle
x^{*}|A_m^{*}|x\rangle=0$. Both conditions induce two sets of
product vectors in $V^{\perp}$, denoted henceforth
as $\mathcal{S}_1$ and $\mathcal{S}_2$.

In the first case, when $x_0=0$, it follows from Eqs.
(\ref{equation}) and (\ref{macierzM}) the vector $\ket{y}$ has to
obey two conditions $y_0=0$ and $\langle
x^{*}|A^{*}_m|y\rangle=\langle
\widetilde{x}^{*}|\widetilde{A}^{*}_m|\widetilde{y}\rangle=0$.
Consequently, the set $\mathcal{S}_1$ consists of product vectors
taking the form
\begin{equation}\label{class1}
(0,\ket{\widetilde{x}})\ot
(0,(\widetilde{A}^{T}_m\ket{\widetilde{x}^{*}})^{\perp})
\end{equation}
with arbitrary $\ket{\widetilde{x}}$
and $(\widetilde{A}^{T}_m\ket{\widetilde{x}^{*}})^{\perp}$
denoting any vector from $(m-2)$-dimensional subspace of
$\mathbbm{C}^{m-1}$ orthogonal to
$\widetilde{A}^{T}_m\ket{\widetilde{x}^{*}}$.

In the second case, when $\langle
x^{*}|A_m^{*}|x\rangle=\langle\widetilde{x}^{*}|\widetilde{A}_m^{*}|\widetilde{x}\rangle=0$,
it clearly follows from (\ref{macierzM}) that
$\ket{y}=(y_0/x_0)\ket{x}$ (we can obviously assume $x_0\neq 0$),
which together with the above condition determines the second set
$\mathcal{S}_2$ of product vectors in $V^{\perp}$.
These are precisely the vectors taking the following form:
\begin{equation}\label{class2}
(x_0,\ket{\widetilde{x}})\ot (x_0,\ket{\widetilde{x}})
\end{equation}
with $\ket{\widetilde{x}}$ obeying
$\langle\widetilde{x}^{*}|\widetilde{A}_m^{*}|\widetilde{x}\rangle=0$.

We are now prepared to show that for an appropriate choice of the
matrix $A_m$, the set $\mathcal{P}_V$ does not span
$\mathcal{H}_m$. For this purpose, we denote by
$\mathcal{S}^{*}_{1}$ and $\mathcal{S}^{*}_{2}$ the partially
conjugated vectors from $\mathcal{S}_{1}$ and $\mathcal{S}_{1}$,
respectively. Then, clearly,
$\mathcal{P}_V=\mathcal{S}_1^{*}\cup\mathcal{S}_2^{*}$.

Since we have still some freedom in the vector $\ket{\Psi_m}$, let
us check under which conditions each set does not span
$\mathcal{H}_m$. First, it is fairly straightforward to see that
$\mathrm{dim}\,\spa\mathcal{S}^{*}_1=(m-1)^2$. For this purpose
let us determine the conditions under which a vector
$\ket{\psi}=\sum_{ij}\alpha_{ij}\ket{ij}\in\mathcal{H}_m$ is
orthogonal to all the elements of $\mathcal{S}^{*}_1$.

At the beginning, observe that due to the form of vectors from
$\mathcal{S}_1$, $\alpha$s with one of the indices being zero are
arbitrary, implying already that $\mathrm{dim}\,\spa
\,\mathcal{S}_1^{*}\leq (m-1)^2$. In order to prove equality, we
show that the remaining $\alpha$s, forming a matrix
$\widetilde{\alpha}$ (see the Appendix), have to be $0$. In terms of
$\widetilde{\alpha}$, the condition for $\ket{\psi}$ to be
orthogonal to $\mathcal{S}_1^{*}$, takes the form
\begin{equation}
 \bra{\widetilde{x}}\widetilde{\alpha}\ket{\widetilde{y}}=0
\end{equation}
for any
$\ket{\widetilde{x}},\ket{\widetilde{y}}\in\mathbbm{C}^{m-1}$
obeying $\langle
\widetilde{x}^{*}|\widetilde{A}_m^{*}|\widetilde{y}\rangle=0$.
This is equivalent to saying that $\ket{\widetilde{y}} \perp
\widetilde{\alpha}^{\dagger}\ket{\widetilde{x}}$ for an arbitrary
$(m-1)$-dimensional $\ket{\widetilde{x}}$ and
$\ket{\widetilde{y}}$ fulfilling
$\ket{\widetilde{y}}\perp\widetilde{A}_m^T\ket{\widetilde{x}^*}$.
By the assumption that $\widetilde{A}$ is of full rank,
$\widetilde{A}_m^T\ket{\widetilde{x}^{*}} \ne 0$ for any
$\ket{\widetilde{x}^*}$. Then, for a fixed $\ket{\widetilde{x}}$,
the vector $\widetilde{\alpha}^{\dagger}\ket{\widetilde{x}}$ is
perpendicular to the $(m-2)$-dimensional subspace
$(\widetilde{A}_m^T\ket{\widetilde{x}^*})^\perp$ in
$\mathbb{C}^{m-1}$, so
$\widetilde{\alpha}^{\dagger}\ket{\widetilde{x}} = k_x
\widetilde{A}_m^T\ket{\widetilde{x}^*}$, where $k_x$ is a scalar
depending on $\ket{\widetilde{x}}$.

Now, one easily checks that for all $\ket{\widetilde{x}}$,
the constant $k_x$ has to be independent of $\ket{\widetilde{x}}$;
let it be denoted by $k$. Consequently, the condition
that $\widetilde{\alpha}$ must obey may be now written as
$\widetilde{\alpha}^{\dagger}\ket{\widetilde{x}} = k
\widetilde{A}_m^T\ket{\widetilde{x}^*}$ for all
$\ket{\widetilde{x}} \in \mathbb{C}^{m-1}$. Let us write the
latter for two vectors, $\ket{\widetilde{x}}$ and
$\mathrm{i}\ket{\widetilde{x}}$, which gives us
\begin{eqnarray}
 \widetilde{\alpha}^{\dagger}\ket{\widetilde{x}} &=& k\widetilde{A}_m^T\ket{\widetilde{x}^*}, \nonumber \\
 \mathrm{i}\widetilde{\alpha}^{\dagger}\ket{\widetilde{x}} &=& -\mathrm{i} k \widetilde{A}_m^T\ket{\widetilde{x}^*} . \nonumber
\end{eqnarray}
It means that for any $\ket{\widetilde{x}} \in
\mathbbm{C}^{m-1}$,
$\widetilde{\alpha}^{\dagger}\ket{\widetilde{x}} = -
\widetilde{\alpha}^{\dagger}\ket{\widetilde{x}}$, and hence
$\widetilde{\alpha}=0$.

Second, the dimension of $\mathrm{span}\,\mathcal{S}^{*}_2$
depends on the matrix $A_m$. This is because one of the conditions
defining $\mathcal{S}_2^{*}$ that $\ket{\widetilde{x}^{*}}\perp
\widetilde{A}_m^{T}\ket{\widetilde{x}}$ can be seen as
a second-order equation for some $x_i$ $(i=1,\ldots,m-1)$. In the
case when the latter has two different solutions, $x_i$ is a
nonlinear function of the remaining $x_j$ $(j\neq i)$ (and also of
the entries of $A_m$) and thus linearly independent of them. In
such case $\mathcal{S}_2^{*}$ spans $\mathcal{H}_m$.

It may happen, however, that
$\langle\widetilde{x}^{*}|\widetilde{A}_m^{*}|\widetilde{x}\rangle=0$
has a single solution with respect to some $x_i$
$(i=1,\ldots,m-1)$ and then
it is just a linear combination of the remaining entries of $\ket{x}$.
As a result, the set $\mathcal{S}^{*}_2$ spans only
$(m-1)^{2}$-dimensional subspace. This is the situation we look
for because it may lead (and, as we will see shortly, does lead) to optimal witnesses without the property of spanning.
For instance, this happens when $\widetilde{A}_m$ is given by
\begin{equation} \label{macierzB}
\widetilde{A}_m = \left( \begin{array}{cccc} 1 & 2 & \dots & 2 \\
0 & 1 & \ddots & \vdots \\ \vdots & \ddots & \ddots & 2 \\ 0 &
\dots & 0 & 1 \end{array} \right).
 \end{equation}
In this case, a direct check shows that one of the conditions
defining the set $\mathcal{S}_2$ reduces now to
\begin{equation}
\langle\widetilde{x}^{*}|\widetilde{A}_m^{*}|\widetilde{x}\rangle=x_1+\ldots+x_{m-1}=0.
\end{equation}

From now on, let us concentrate on this particular case [i.e., when
$\widetilde{A}_m$ is given by Eq. (\ref{macierzB})] and show that
the completely entangled subspace $V$ defined in this way does not
have the property of spanning, that is, $\mathcal{P}_V$ does not
span $\mathcal{H}_m$. To this end, we determine
$(\spa\,\mathcal{P}_V)^{\perp}$. Let us take again the vector
$\ket{\psi}=\sum_{i,j=0}^{m-1}\alpha_{ij}\ket{ij}$ from
$\mathcal{H}_m$. We already know that it is orthogonal to
$\mathcal{S}^{*}_1$ iff $\alpha_{ij}=0$ for $i,j=1,\ldots,m-1$.
Then, the resulting $\ket{\psi}$ is orthogonal to
$\mathcal{S}^{*}_2$ iff $\alpha_{00}=0$,
$\alpha_{0i}=\alpha_{0,m-1}$, and $\alpha_{i0}=\alpha_{m-1,0}$
with $i=1,\ldots,m-2$. As a consequence, the orthogonal complement
of $\mathrm{span}\,\mathcal{P}_V$ is a two-dimensional subspace of
$\mathcal{H}_m$ spanned by the vectors
\begin{eqnarray}
\ket{\phi_1}=\ket{0}\ket{\omega},\qquad
\ket{\phi_2}=\ket{\omega}\ket{0},
\end{eqnarray}
where $\ket{\omega}=\ket{1}+\ldots+\ket{m-1}$.

\subsection{The case of arbitrary dimensions}

Before going into detail, let us comment
on the notation we use throughout this section. Given a
vector $\ket{z}$ from $\mathbbm{C}^{n}$, we often
decompose it as $\ket{z}=\ket{z'}\oplus\ket{z''}$ with
$\ket{z'}\in\spa\{\ket{0},\ldots,\ket{m-1}\}$ and the rest
$\ket{z''}$.

In this section, we extend the already defined $m$-dimensional CES
$V$ of $\mathcal{H}_m$ to an $n$-dimensional CES $\overline{V}$ in
the arbitrary dimensional Hilbert space $\mathcal{H}_{m,n}$ with
$m,n\geq 3$. For simplicity, let us assume that $n>m$. We
prove that the subspace $\overline{V}$ does not have the
property of spanning.

First, let us take the vectors $\ket{\Psi_i}$ $(i=1,\ldots,m)$
(with $\ket{\Psi_m}$ defined in Eq. (\ref{macierzB}), embed them
in $\mathbbm{C}^{m}\ot\mathbbm{C}^n$ $(n>m)$, and then supply with
the following $m-n$ vectors:
\begin{equation}\label{PsiII}
\ket{\Psi_i}=\ket{1}\ket{i-2}-\ket{2}\ket{i-1} \qquad
(i=m+1,\ldots,n).
\end{equation}
Clearly, $\overline{V}=\spa\{\ket{\Psi_i}\}_{i=1}^{n}$ is a
$n$-dimensional subspace in $\mathcal{H}_{m,n}$, which
does not contain any product vector (see the Appendix for the detailed
proof).

As before, looking for the product vectors $\ket{x,y}$
$(\ket{x}\in\mathbbm{C}^m,\ket{y}\in\mathbbm{C}^n)$ orthogonal to
$\overline{V}$, we arrive at the system of linear equations
[similar to the one given in Eq. \ref{system}], which can be stated
as
\begin{equation}\label{systemII}
N(x)\ket{y}=0,
\end{equation}
with the $n\times n$ matrix $N(x)$ given by
\begin{equation}\label{systemMac}
N(x)=\left(
\begin{array}{cccc|ccccc}
& & & & & & & & \\
&  &\hspace{-0.5cm}M(x)&  & & & \mathbf{0}_{n-m}& & \\
& & & & & & & & \\
\hline
0 & \ldots & 0 & x_0& -x_1 & 0       & 0     & \ldots & 0\\
0 & \ldots &  \ldots  & 0 &        x_0     & -x_1   & 0 &  \ldots & 0\\
0 & \ldots & \ldots & 0   & 0   & x_0 & -x_1 & \ldots & 0\\
\vdots & \vdots & \vdots & \vdots & \vdots & \vdots & \ddots & \ddots & \vdots\\
0 & \ldots & \ldots  & 0 & 0 & \ldots & 0 & x_0 & -x_1
\end{array}
\right),
\end{equation}
$M(x)$ being the $m\times m$ matrix already introduced in Eq.
(\ref{macierzM}), and $\textbf{0}_{n-m}$ being the $(n-m)\times (n-m)$
zero matrix.

The above system has nontrivial solutions iff $\det N(x)=0$,
meaning that either $\det M(x)=0$ or $x_1=0$. We
already know that the first condition holds iff either $x_0=0$ or
$x_1+\cdots+x_{m-1}=0$, giving us, as before, two sets of product
vectors orthogonal to $\overline{V}$ (denoted respectively by
$\mathcal{\overline{S}}_1$ and $\mathcal{\overline{S}}_2$).
Additionally, however, we have a third set of vectors
$\mathcal{\overline{S}}_3$ corresponding to the third solution
$x_1=0$ of the above determinantal equation. Let us now determine
these sets.

The first one, $\overline{\mathcal{S}}_1$, we get by putting
$x_0=0$ in Eq. (\ref{systemMac}). In this case, the matrix $N(x)$
has a simple block-diagonal form, that is, $N(x)=M(x)\oplus
(-x_1)\mathbbm{1}_{n-m}$, with $\mathbbm{1}_{d}$ denoting the
$d\times d$ identity matrix. Clearly, Eq. (\ref{systemII}) implies
that $\ket{y}$ has to satisfy $-x_1\ket{y''}=0$ and
$M(x)\ket{y'}=0$. While the former condition immediately gives
$\ket{y''}=0$, the latter one, together with $x_0=0$, implies as
before that $y_0=0$ and that $\ket{y'}$ has to be orthogonal to
$A_m^{T}\ket{x^{*}}$ or, in other words, has to obey the equation
\begin{equation}\label{condition}
\sum_{i=1}^{m-1}x_iy_i+2\sum_{i<j=1}^{m-1}x_j y_k=0.
\end{equation}
In conclusion, $\overline{\mathcal{S}}_1$ consists of vectors
\begin{equation}
(0,x_1,\ldots,x_{m-1})\ot [(0,y_{1},\ldots,y_{m-1})\oplus 0_{n-m}
],
\end{equation}
where all $x$s and $y$s have to obey the condition
(\ref{condition}) and $0_{n-m}$ stands for $(n-m)$-dimensional
zero vector. Let us notice that this is the same class as
$\mathcal{S}_1$ (see Sec. \ref{equal}) but is embedded in a
larger Hilbert space $\mathcal{H}_{m,n}$.

In the second case, when $x_1+\cdots+x_{m-1}=0$, Eqs.
(\ref{systemII}) and (\ref{systemMac}) imply that $M(x)\ket{y'}=0$
and $y_{i}=(x_0/x_1)y_{i-1}$ for $i=m,\ldots,n-1$. From the first
condition we simply get $\ket{y'}=(y_0/x_0)\ket{x}$. The remaining
equations can be rewritten as $y_i=(x_0/x_1)^{i-m+1}y_{m-1}$ and
then, taking into account that $y_{m-1}=(y_0/x_0)x_{m-1}$, as
$y_i=(y_0/x_0)(x_0/x_1)^{i-m+1}x_{m-1}$ with $i=m,\ldots,n-1$.
Consequently, the second set $\overline{\mathcal{S}}_3$ has the
following elements:
\begin{equation}
\ket{x}\ot [\ket{x}\oplus x_{m-1}(t,t^2,\ldots,t^{n-m}) ],
\end{equation}
where $t=y_0/x_0$ and $x$s have to satisfy $x_1+\cdots+x_{m-1}=0$.

The third set $\overline{\mathcal{S}}_1$ consists of product
vectors solving the system (\ref{systemII}) in the case when
$x_1=0$. From Eqs. (\ref{systemII}) and (\ref{systemMac}), it
follows that $\ket{y}$ has to obey $M(x)\ket{y'}=0$ and $x_0
y_i=0$ for $i=m-1,\ldots,n-2$. While the latter conditions
immediately imply that $y_i=0$ for $i=m-1,\ldots,n-2$, the former,
together with the initial condition $x_1=0$, yields $y_1=0$,
$\ket{y'}=(y_0/x_0)\ket{x}$, and $x_2+\cdots+x_{m-2}=0$. Taking
all these conditions together, we see that the product vectors
from $\overline{\mathcal{S}}_1$ take the form
\begin{equation}
\ket{x}\ot [(y_0/x_0)\ket{x}\oplus (0,\ldots,0,y_{n-1})],
\end{equation}
with $\ket{x}$ given by
\begin{equation}
\ket{x}=(x_0,0,x_2,\ldots,x_{m-2},0),
\end{equation}
where $x_2+\cdots+x_{m-2}=0$.

Having determined all the product vectors in $V^{\perp}$, we can
now show that $\mathcal{P}_V$ does not span $\mathcal{H}_{m,n}$.
Recall that $\mathcal{P}_V$ consists of partial conjugations of
elements of the sets $\mathcal{S}_i$. Denoting partially
conjugated vectors from $\overline{\mathcal{S}}_i$ by
$\overline{\mathcal{S}}_i^{*}$ $(i=1,2,3)$, we have
$\mathcal{P}_V=\overline{\mathcal{S}}_1^{*}\cup\overline{\mathcal{S}}_2^{*}\cup\overline{\mathcal{S}}_3^{*}$.

First, let us notice that
$\spa\,\mathcal{P}_V=\spa\,\overline{\mathcal{S}}_1^{*}\cup\overline{\mathcal{S}}_2^{*}$.
Second, let us take a vector $\ket{\psi}\in\mathcal{H}_{m,n}$ with
entries in the standard basis denoted by $\alpha_{ij}$. Short
algebra shows that it is orthogonal to
$\overline{\mathcal{S}}_1^{*}$ iff $\alpha_{ij}=0$
$(i,j=1,\ldots,m-1)$. Then, $\ket{\psi}$ is orthogonal to
$\overline{\mathcal{S}}_2^{*}$ iff the following set of condition
holds
\begin{eqnarray}
\begin{array}{ll}
\alpha_{00}=0, & \\
\alpha_{0j}=0 & (j=m,\ldots,n-1),\\
\alpha_{i0}=\alpha_{10} & (i=2,\ldots,m-1),\\
\alpha_{0j}=\alpha_{01} & (j=2,\ldots,m-1),\\
\alpha_{ij}=\alpha_{1j} & (i=2,\ldots,m-1;j=m,\ldots,n-1).
\end{array}
\end{eqnarray}
From this analysis, it clearly follows that $\mathcal{P}_V$ does not span $\mathcal{H}_{m,n}$ and the $(n-m+2)$-dimensional subspace $K=(\spa\,\mathcal{P}_V)^{\perp}$
is spanned by the vectors $\ket{\phi_1}$, $\ket{\phi_2}$, and
$\ket{\phi_{j-m+3}}=\ket{\omega}\ket{j}$ $(j=m,\ldots,n-1)$.

\subsection{Optimal decomposable witnesses}

What remains to be proven is that the subspace $\overline{V}$
admits optimal decomposable witnesses, that is, one is able to find a
positive $Q$ supported on $\overline{V}$ of rank $n$ such that
$Q^{\Gamma}$ is a optimal decomposable EW. As we will see shortly,
this can be achieved by taking $Q$ of the form
\begin{equation}\label{Q}
Q=\sum_{i=1}^n \lambda_i\proj{\Psi_i}
\end{equation}
with $\lambda_i>0$ $(i=1,\ldots,n)$. Clearly, by definition, $Q$
is a positive matrix supported on $\overline{V}$ of rank $n$, and
moreover, it is straightforward to see that it is NPT for any
choice of $\lambda$s, thus giving rise to proper DEWs. Notice also
that here $\mathcal{P}_W$ of $W=Q^{\Gamma}$ is exactly the same as
$\mathcal{P}_{\overline{V}}$.

According to what was said in Sec. \ref{Preliminaries}, we need to prove that the operator $\widetilde{W}(\epsilon,P)=Q^{\Gamma}-\epsilon P$ is no longer a witness for any $\epsilon>0$, with $P$ being positive matrices obeying $P\perp \mathcal{P}_{\overline{V}}$.
The latter are the positive operators supported on
$(n-m+2)$-dimensional subspace $\mathcal{K}=(\spa
\mathcal{P}_V)^{\perp}=\spa\{\ket{\phi_i}\}_{i=1}^{n-m+2}$.
Let us prove our statement for the case when $P$ is just a
one-dimensional projector onto a general vector from
$\mathcal{K}$, that is,
\begin{equation}
\ket{\varphi}=\sum_{i=1}^{n-m+2}a_i \ket{\phi_i}\qquad
(a_i\in\mathbbm{C}).
\end{equation}
Then, clearly, the proof will follow for an arbitrary $P\geq 0$
supported on $K$.

Denoting by $P_{\varphi}$ the projector onto $\ket{\varphi}$, one
checks that for the product vector $\ket{u}\ot \ket{v^{*}}$, where
$\ket{v}=\ket{u}\oplus u_0 1_{n-m}$ and $u_0=u_1=u_{m-1}$, the
following holds:
\begin{eqnarray}\label{rownanie}
&&\langle u,v^{*}|Q^{\Gamma}-\epsilon P_{\varphi}|u,v^{*}\rangle=\langle u,v|Q|u,v\rangle-\epsilon|\langle \varphi|u,v^{*}\rangle|^{2}\nonumber\\
&&=\lambda_{m}|\langle\Psi_m |u,u\rangle|^{2}-\epsilon|\langle \varphi|u,v^{*}\rangle|^{2}\nonumber\\
&&=\lambda_{m}|\overline{u}|^{4}-\epsilon|a_{1}u_0\overline{u}^{*}
+(a_2+a_3+\ldots+a_{n-m+2})u_{0}^{*}\overline{u}|^{2},\nonumber\\
\end{eqnarray}
where we denoted $\overline{u}=u_1+\cdots+u_{m-1}$. The second
equality is a consequence of the fact that $\ket{u}\ket{v}$ is
orthogonal to $\ket{\Psi_i}$ $(i=1,\ldots,m-1)$ and $\ket{\Phi_i}$
$(i=m,\dots,n-1)$, while the third one stems from the fact that
$\langle\Psi_m|u,u\rangle=(u_1+\cdots+u_{m-1})^{2}=\overline{u}^2$.

For simplicity, let us now put $u_0=1$ and denote
$a=a_2+\dots+a_{n-m+2}$. Then, Eq. (\ref{rownanie}) can be
rewritten as
\begin{eqnarray}\label{obliczenia2}
&\langle u,v^{*}|Q^{\Gamma}-\epsilon P_{\varphi}|u,v^{*}\rangle=\lambda_{m}|\eta|^4-\epsilon|a_1 \eta+a\eta^{*}|^2\nonumber\\
&=|\eta|^{2}\left(\lambda_{m}|\eta|^2-\epsilon
\left|a_1+a\mathrm{e}^{-2\mathrm{i}\delta}\right|^2\right),
\end{eqnarray}
where the second equality follows from the fact that we can always
write $\eta=|\eta|\mathrm{e}^{\mathrm{i}\delta}$. Clearly, we can
always choose $\delta$ such that
$|a_1+a\mathrm{e}^{-2\mathrm{i}\delta}|>0$ and then, taking
sufficiently small $|\eta|$, we can make $\langle
u,v^{*}|Q^{\Gamma}-\epsilon P_{\varphi}|u,v^{*}\rangle$ negative
for any $\epsilon>0$.

Eventually, let us observe that the same reasoning applied when
$P$ is any positive matrix supported on $\mathcal{K}$.

Concluding $Q$ given by Eq. (\ref{Q}) gives rise to an optimal
decomposable witness such that the corresponding $P_W$ does not
span $\mathcal{H}_{m,n}$.

\section{Further results on the general optimal decomposable witnesses}
\label{Further}

Let us now ask whether and how the results obtained in Ref.
\cite{nasza} for qubit-qunit witnesses can be generalized to DEWs
acting on $\mathcal{H}_{m,n}$ with $m,n\geq 3$.
Following Ref. \cite{nasza} and the structure of equations
defining product vectors in an orthogonal complement of some CES,
we surmise that for any CES $V$ of $\mathcal{H}_{m,n}$ with $\mathrm{dim}
V\leq n-1$ the corresponding $\mathcal{P}_V$ spans
$\mathcal{H}_{m,n}$. The main purpose of this section is to
comment on this issue.

First, we show that under some assumption, which is
generically obeyed and conjectured to hold always, all CESs of
$\mathcal{H}_n$ of dimension $n-1$ have the property of spanning.
Then, we prove that any $r$-dimensional CES is a subspace of some
$(r+1)$-dimensional CES. Consequently, provided the above
conjecture holds, all CESs in $\mathcal{H}_n$ of dimension $n-1$ or
less have the property of spanning.

Then, we prove that the mentioned assumption is always fulfilled for CESs supported in $\mathcal{H}_4$ of dimension 3.
Therefore, due to the above fact, any CES supported in
$\mathcal{H}_4$ of dimension less or equal to 3 has the
property of spanning, and hence for all witnesses (\ref{form}) in
$\mathbbm{M}_4$ with $r(Q)\leq 3$, the statements (i) and (ii)
(see Sec. \ref{Preliminaries}) become equivalences.

To be more precise, let $V\subset \mathcal{H}_n$ be an
$(n-1)$-dimensional CES spanned by $\ket{\Psi_i}$
$(i=1,\ldots,k)$. Product vectors in $V^{\perp}$ must obey the set
of equations $\langle \Psi_i|x,y\rangle=0$ $(i=1,\ldots,n-1)$,
which, as we already know, can be rewritten as
\begin{equation}\label{systemEQ}
B(x)\ket{y}=0,
\end{equation}
with the $\ket{x}$-dependent $(n-1)\times n$ matrix $B(x)$. For
further benefit, let us denote by $\Pi_V$ the projector onto $V$
and by $\Pi_V(x)$ a local projection of $\Pi_V$ onto
$\ket{x}\in\mathbbm{C}^n$, that is,
$\Pi_V(x)=\Tr[(\proj{x}\ot\mathbbm{1})\Pi_V]$. Analogously,
$\ket{\Psi(x)}$ stands for the local projection of some composite
vector $\ket{\Psi}\in\mathcal{H}_{n}$ onto
$\ket{x}\in\mathbbm{C}^n$. In both cases, for concreteness, we
always project at the first subsystem.

Now we are prepared to prove the following theorem.
\begin{thm}
 Consider an $(n-1)$-dimensional subspace $V \subset \mathcal{H}_n$ possessing the following properties:
 \begin{enumerate}
  \item Local projection of $\Pi_V$ onto at least one vector $\ket{x}\in\mathbbm{C}^{n}$ gives a full-rank density matrix of the other subsystem.

  \item The subspace $V$ cannot be embedded in any $V_1 \otimes V_2$, where $V_i$ or $V_2$ is a proper subspace of $\mathbb{C}^n$.
 \end{enumerate}
 Then $\mathcal{P}_V$ spans the whole $\mathcal{H}_n$.
\end{thm}

\noindent\textbf{Proof.} Assume that $V$ is spanned by the the
following orthonormal vectors:
\begin{equation} \label{Psis}
\ket{\Psi_i} = \sum_{k,l=0}^{n-1} a_{kl}^{(i)} \ket{k}\ket{l}
\quad (i=0, \dots, n-1).
\end{equation}
As before, $A_i$ $(i=0,\ldots,n-1)$ are square matrices
corresponding to $\ket{\Psi_i}$, that is, matrices formed from the
numbers $a_{kl}^{(i)}$.

Product vectors in $V^{\perp}$ are solutions to the system of
equations (\ref{systemEQ}). Assume now that for some
$\ket{x}\in\mathbbm{C}^{n}$ the matrix $B(x)$ is of full rank, and
denote by $M_i(x)$ $(n=0,\ldots,n-1)$ the matrices obtained by
removing the $i$th column from $B(x)$. Then this system has a
unique solution (and the corresponding product vector in
$V^{\perp}$) given by $\ket{y(x)}=[y_0(x),\ldots,y_{n-1}(x)]$ with
$y_i(x)$ being the determinant of $M_i(x)$, that is,
 \begin{eqnarray} \label{y_i}
y_i(x)=\det M_i(x)\qquad (i=0,\ldots,n-1).
 \end{eqnarray}

Now, let us notice that $B(x)$ is of full rank iff $\Pi_V(x)$ is.
This is because
\begin{equation}
\Pi_V(x)=\sum_{i}\proj{\Psi_i(x)},
\end{equation}
and, clearly, $\Pi_V(x)$ is of full rank iff all the vectors
$\ket{\Psi_i(x)}$ $(i=1,\ldots,n-1)$ are linearly independent.
Since $\ket{\Psi_i(x)}$ are just rows of $B(x)$, we infer that
$B(x)$ is of full rank iff $\Pi_V(x)$ is.

The first assumption, together with the above fact, tell us that
there exists a vector $\ket{x}\in\mathbbm{C}^n$ such that $B(x)$ is
of full rank. Let $\ket{x_0}$ denote such a vector.
Consequently one of the matrices $M_i(x)$ has nonvanishing
determinant at $\ket{x_0}$. On the other hand, by the very
definition, $y_i(x)$ is a homogenous polynomial in coordinates of
$\ket{x}$. Since then $y_i(x)$ does not vanish at at least one
$x$, it cannot be identically equal zero. As, moreover, the
equation $y_i(x)=0$ has solutions only in the set of Lebegue's
measure zero, the matrix $B(x)$ is of full rank for almost all
$\ket{x}\in\mathbbm{C}^{n}$. In other words, the system
(\ref{systemEQ}) has a unique solution for almost all $\ket{x}$s
given by Eqs. (\ref{y_i}). For the remaining $\ket{x}$s, it has at
least a two-dimensional subspace of solutions; however, in such
cases Eqs. (\ref{y_i}) give us a zero vector.
Nevertheless, the set of product vectors
$\ket{x}\ket{y(x)}$ with $\ket{y(x)}$ given by Eqs. (\ref{y_i}) is
enough to span, after being partially conjugated, the Hilbert
space $\mathcal{H}_n$. For convenience, let us denote these
solutions by $\mathcal{C}$ and their partial conjugations by
$\mathcal{C}^{*}$.

Notice that the above reasoning holds when we exchange the
subsystems, that is, for almost all $\ket{y}\in\mathbbm{C}^n$ there
exist $\ket{x}\in\mathbbm{C}^n$ such that $y$ is the unique (up to
a scalar) solution of Eq. (\ref{systemEQ}).

Because the polynomial formulas (\ref{y_i}) for coordinates of the
solution of Eq. (\ref{systemEQ}) produce almost all vectors $\ket{y}$
from $\mathbbm{C}^n$, the polynomials (\ref{y_i}) are linearly
independent.

It remains to prove that the vectors from $\mathcal{C}^{*}$,
(i.e., vectors $\ket{x}\ot \ket{y(x^{*})}$) span $\mathbb{C}^n
\otimes \mathbb{C}^n$. To this end, observe that $\ket{y}$ is
defined by $n$ linearly independent polynomials of variables
$x_0^*, \dots, x_{n-1}^*$. Moreover, they are linearly independent
of the polynomials $x_0, \dots, x_{n-1}$ (no $x_i^*$ can be
achieved by a combination of $x_i$s so also no $y_i^*$). It
implies, that the coordinates of $\ket{x} \otimes \ket{y(x^{*})}$
form a set of $n^2$ linearly independent polynomials, so the set
$\{ \ket{x} \otimes \ket{y(x^{*})}:\ \ket{x} \in \mathbb{C}^n \ \land \ \mathrm{rank} B(x) = n-1 \}$
spans the whole $\mathbb{C}^n \otimes \mathbb{C}^n$. $\square$\\

\noindent\textbf{Remark 1.} Any subspace $W$ in $\mathbb{C}^n
\otimes \mathbb{C}^n$ of dimension less than $n-1$, which can be
extended to $(n-1)$-dimensional subspace (which we show to
be the case in Lemma 2), satisfying the assumptions of the
theorem, inherits the property of spanning from $V$. Indeed, all
product vectors orthogonal to $V$ are orthogonal to $W$, so
already a subset of product vectors orthogonal to
$W$ spans the whole $\mathbb{C}^n \otimes \mathbb{C}^n$ after partial conjugation.\\

\noindent\textbf{Remark 2.} It should be noticed that for any CES
in $\mathcal{H}_{2,n}$ the first assumption of the theorem is
always obeyed. In order to see it explicitly, let us write the
orthonormal vectors spanning such CES as
\begin{equation}
\ket{\Psi_i}=\ket{0}\ket{\Psi_0^{(i)}}+\ket{1}\ket{\Psi_{1}^{(i)}}\qquad
(i=1,\ldots,\mathrm{dim}V)
\end{equation}
Assume now that in this case the local projection of $\Pi_V$ is
rank deficient for any $\ket{x}\in\mathbbm{C}^2$. Taking the
vector $\ket{x}=\ket{0}$, the matrix $\Pi_V(x)$ is just a sum of
projections onto $\ket{\Psi_0^{(i)}}$. Since here $\Pi_V(x)$ is
not of full rank, the vectors $\ket{\Psi_0^{(i)}}$ must be
linearly dependent. This, however, immediately implies that there
is a product vector in $V$ which contradicts the assumption that $V$ is CES.

Now, let us show that for any three-dimensional CES supported in
$\mathcal{H}_4$, the first assumption of the theorem is satisfied.\\

\noindent \textbf{Lemma 1.} {\it Let $V$ be a three-dimensional
CES supported in $\mathbbm{C}^4\ot \mathbbm{C}^4$. Then there
exist $\ket{x}\in\mathbbm{C}^4$
such that the matrix $B(x)$ is of rank 3.} \\

\noindent\textbf{Proof.} Assume in contrary that for any
$\ket{x}\in\mathbbm{C}^4$ the matrix $B(x)$ is of rank 2 or
lower. Notice that rows of $B(x)$ are just vectors $\ket{\Psi_i}$
$(i=1,2,3)$ spanning $V$ projected locally on $\ket{x}$
(henceforward denoted by $\ket{\Psi_i(x)}$). Also, in what follows,
we denote by $\ket{z'}$ and $\ket{z''}$ the
$\ket{z}\in\mathbbm{C}^4$ projected onto
$\mathrm{span}\{\ket{1},\ket{2},\ket{3}\}$ and
$\mathrm{span}\{\ket{0},\ket{2},\ket{3}\}$, respectively.

Let us start the proof by noting that in
$\mathbbm{C}^4\ot\mathbbm{C}^4$ the largest subspace with all
vectors having Schmidt rank 4 is one-dimensional (see, e.g.,
Refs. \cite{Parthasarathy,CubittJMP08,WalgateJPA08}). Therefore,
we can assume that one of the vectors spanning our CES $V$, say
$\ket{\Psi_1}$, has Schmidt rank either 2 or 3. As the proof
is straightforward but tedious, in what follows we consider
only the first case; however, analogous reasoning works also for
the case of Schmidt rank 3.

By local unitary operations, we can always bring $\ket{\Psi_1}$ to
$\ket{\Psi_1}=\ket{00}+\ket{11}$, while the remaining two vectors
take the general forms
\begin{eqnarray}\label{wektory1}
\ket{\Psi_2}&=&\ket{0}\ket{\psi_0}+\ket{1}\ket{\psi_1}+\ket{2}\ket{\psi_2}+\ket{3}\ket{\psi_3}\nonumber\\
\ket{\Psi_3}&=&\ket{0}\ket{\varphi_0}+\ket{1}\ket{\varphi_1}+\ket{2}\ket{\varphi_2}+\ket{3}\ket{\varphi_3}
\end{eqnarray}
Our proving strategy is that we consider vectors $\ket{\Psi_i(x)}$
$(i=1,2,3)$ for some particular subsets of vectors of
$\mathbbm{C}^4$. Then, demanding that all the $3\times3$
submatrices of the $3\times 4$ matrix $B(x)$ 
have zero determinant, we show that $V$ is not either a CES or
supported in $\mathcal{H}_4$.

We start by noting that the above procedure for $\ket{x}=\ket{0}$
and $\ket{x}=\ket{1}$ leads us to a conclusion that the sets
$\{\ket{0},\ket{\psi_0},\ket{\varphi_0}\}$ and
$\{\ket{1},\ket{\psi_1},\ket{\varphi_1}\}$ must be linearly
dependent. This means that by taking appropriate linear
combinations of $\ket{\Psi_i}$ $(i=1,2,3)$, the vectors spanning
$V$ can be expressed as
\begin{eqnarray}\label{wektory2}
\ket{\Psi_1}&=&\ket{00}+\ket{11}\nonumber\\
\ket{\Psi_2}&=&\ket{0}\ket{\psi_0}+\ket{2}\ket{\psi_2}+\ket{3}\ket{\psi_3}\nonumber\\
\ket{\Psi_3}&=&\ket{1}\ket{\varphi_1}+\ket{2}\ket{\varphi_2}+\ket{3}\ket{\varphi_3},
\end{eqnarray}
where, in general, vectors $\ket{\psi_i}$ and $\ket{\varphi_i}$
are different than the ones appearing in Eq. (\ref{wektory1}).
However, for convenience, we do not change the notation.

In what follows, we split the proof into two cases, that is, when both
vectors $\ket{\psi_0}$ and $\ket{\varphi_1}$ are nonzero and one
of them vanishes.

{\it The case of nonzero $\ket{\psi_0}$ and $\ket{\varphi_1}$.} By
choosing $\ket{x}=(1,\alpha,0,0)$ $(\alpha\in\mathbbm{C})$ and
demanding that the corresponding matrix $B(x)$ is of rank at most
2 (i.e., that all the $3\times 3$ determinants of this matrix
vanish), we get the conditions that either
$\ket{\psi_0}\parallel\ket{\varphi_1}$ or
$\ket{\psi_0},\ket{\varphi_1}\in\mathrm{span}\{\ket{0},\ket{1}\}$.

In both cases, we can assume that either $\ket{\psi_0}\nparallel
\ket{0}$ or $\ket{\varphi_1}\nparallel\ket{1}$ because otherwise
one can remove $\ket{\psi_0}$ or $\ket{\varphi_1}$ from the
vectors (\ref{wektory2}) by taking appropriate linear combinations
of $\ket{\Psi_i}$ $(i=1,2,3)$, leading us to the case considered
below. Let us then assume that $\ket{\psi_0}\nparallel \ket{0}$,
which in particular means that $\ket{\psi_0'}\neq 0$.

Now, we project onto $\ket{x}=(1,0,\alpha,\beta)$
$(\alpha\in\mathbbm{C})$, which allows us to conclude that
$\ket{\psi_0'}$ has to be parallel to all the vectors of
$\ket{\psi_i'}$ and $\ket{\varphi_i'}$ $(i=2,3)$ which are
nonzero, unless $\ket{\varphi_2'}=0$ and $\ket{\varphi_3'}=0$. In
the former case, one immediately finds that $V$ is not supported in
$\mathcal{H}_1$. This is because either $\ket{\psi_0}$ is
proportional to $\ket{\varphi_1}$, meaning that
$\ket{\psi_0'}\parallel\ket{\varphi_1'}
\parallel\ket{\psi_2'}\parallel\ket{\varphi_2'}\parallel\ket{\psi_3'}\parallel\ket{\varphi_3'}$
and therefore
$V\subseteq\mathbbm{C}^4\ot\mathrm{span}\{\ket{0},\ket{1},\ket{\psi_0'}\}$,
or when
$\ket{\psi_0},\ket{\varphi_1}\in\mathbbm{span}\{\ket{0},\ket{1}\}$,
the vectors $\ket{\psi_0'}$ and $\ket{\psi_i'}$ and
$\ket{\varphi_i}$ $(i=2,3)$ are parallel and hence
$V\subseteq\mathbbm{C}^{4}\ot\mathbbm{C}^2$.

In the case when $\ket{\varphi_2'}=0$ and $\ket{\varphi_3'}=0$, we
notice that it cannot happen that
$\ket{\varphi_2}=\ket{\varphi_3}=0$ because in such case there exists a product vector $\ket{1}\ket{\varphi_1}$ in $V$. Therefore, one of
the vectors $\ket{\varphi_2}$ or $\ket{\varphi_3}$ is proportional
to $\ket{0}$. We project further onto $\ket{x}=(0,1,\alpha,\beta)$
$(\alpha,\beta\in\mathbbm{C})$, which, in the same way as before,
leads us to the additional conditions that
$\ket{\psi_2''}\parallel\ket{\psi_3''}\ket{0}$, and so
$\ket{\psi_2},\ket{\psi_3}\in\mathrm{span}\{\ket{0},\ket{1}\}$.
Then all vectors $\ket{\psi_i}$ and $\ket{\varphi_i}$
$(i=2,3)$ live in $\mathrm{span}\{\ket{0},\ket{1}\}$.
Consequently, irrespectively of whether $\ket{\psi_0}$ is
proportional to $\ket{\varphi_1}$ or
$\ket{\psi_0},\ket{\varphi_1}\in\{\ket{0},\ket{1}\}$, $V$ is not
supported in $\mathcal{H}_4$.

{\it The case when one of the vectors
$\ket{\psi_0},\ket{\varphi_1}$ vanishes.} Without any loss of
generality we can assume that $\ket{\varphi_1}=0$. This
immediately implies that one of the vectors $\ket{\varphi_2'}$ or
$\ket{\varphi_3'}$ is nonzero because otherwise
$\ket{\Psi_3}=\ket{\omega}\ket{0}$ for some
$\ket{\omega}\in\mathbbm{C}^4$, meaning that there is a product
vector in $V$.

We now project locally $\ket{\Psi_i}$ $(i=1,2,3)$ onto
$\ket{x}=(1,0,\alpha,\beta)$ $(\alpha,\beta\in\mathbbm{C})$.
Demanding that the resulting matrix $B(x)$ is of rank less than or
equal to 2, we get the condition that if $\ket{\psi_0'}$ is
nonzero the vectors $\ket{\psi_i'}$ $(i=0,2,3)$ and
$\ket{\varphi_i'}$ $(i=2,3)$ are parallel (excluding the ones that
possibly vanish). As a result, $V$ is a subspace supported on
$\mathbbm{C}^4\ot\mathrm{span}\{\ket{0},\ket{1},\ket{\psi_0'}\}$.

If, on the other hand, $\ket{\psi_0'}=0$
($\ket{\psi_0'}\parallel\ket{0}$), one gets the conditions that
either $\ket{\varphi_2'}=\eta\ket{\psi_2'}$ and
$\ket{\varphi_3'}=\eta\ket{\psi_3'}$ or
$\ket{\psi_2'}\parallel\ket{\varphi_2'}\parallel\ket{\psi_3'}\parallel\ket{\varphi_3'}$
(again excluding the vectors that vanish). In the first of these
cases, we can subtract $\ket{\Psi_3}$ from $\ket{\Psi_2}$
obtaining a product vector in $V$, while in the second case, again
$V$ is supported on a subspace
$\mathbbm{C}^4\ot\mathrm{span}\{\ket{0},\ket{1},\ket{\psi_2'}\}$,
leading to a contradiction with one of the assumptions.

Now let us consider the case when one of the vectors is of Schmidt rank 3 and there is no vector of Schmidt rank 2 in the CES $V$. The vectors spanning $V$ can be written in this case as
\begin{eqnarray}\label{wektorySR3}
\ket{\Psi_1}&=&\ket{00}+\ket{11}+\ket{22},\nonumber\\
\ket{\Psi_2}&=&\ket{0}\ket{\psi_0}+\ket{2}\ket{\psi_2}+\ket{3}\ket{\psi_3},\nonumber\\
\ket{\Psi_3}&=&\ket{1}\ket{\varphi_1}+\ket{2}\ket{\varphi_2}+\ket{3}\ket{\varphi_3},
\end{eqnarray}
where the sets $\{\ket{\psi_0}, \ket{\psi_2}, \ket{\psi_3}\}$ and $\{\ket{\varphi_1}, \ket{\varphi_2}, \ket{\varphi_3}\}$ are linearly independent (otherwise the Schmidt rank of $\ket{\Psi_2}$ or $\ket{\Psi_3}$ would be less than 3). In particular, none of these six vectors can be zero.

Choosing $\ket{x}=(1,0,0,\alpha)$ and $\ket{x}=(0,1,0,\alpha)$ $(\alpha\in\mathbb{C})$, one gets that $\ket{\psi_0'}\parallel\ket{\psi_3'}\parallel\ket{\varphi_3'}$ or $\ket{\varphi_3'}=0$
and $\ket{\varphi_1''}\parallel\ket{\psi_3''}\parallel\ket{\varphi_3''} $ or $\ket{\psi_3''}=0$.
This gives us four cases: (i) $\ket{\psi_0'}\parallel\ket{\psi_3'}\parallel\ket{\varphi_3'}$ and
$\ket{\varphi_1''}\parallel\ket{\psi_3''}\parallel\ket{\varphi_3''}$,
(ii) $\ket{\psi_0'}\parallel\ket{\psi_3'}\parallel\ket{\varphi_3'}$
and $\ket{\psi_3''}=0$, (iii) $\ket{\varphi_1''}\parallel\ket{\psi_3''}\parallel\ket{\varphi_3''} $ and $\ket{\varphi_3'}=0$, and (iv) $\ket{\varphi_3'}=0$ and  $\ket{\psi_3''}=0$ (i.e., $\ket{\varphi_3}\parallel \ket{0}$ and $\ket{\psi_3}\parallel \ket{1}$).

In the first case, we project onto $(1,\alpha,0,0)$ $(\alpha\in\mathbbm{C})$, which gives us additional conditions that either $\ket{\psi_0}\parallel \ket{\varphi_1}$ or $\ket{\psi_0},\ket{\varphi_1}\in\mathrm{span}\{\ket{0},\ket{1}\}$.
In the first of these two possibilities, we can additionally assume that $\ket{\psi_0'}\nparallel\ket{1}$ because otherwise we fall into the second possibility. This implies that in particular $\ket{\psi_0}\parallel\ket{\psi_3}$, meaning that $\ket{\Psi_2}$ is of Schmidt rank 2.

In the case when $\ket{\psi_0},\ket{\varphi_1}\in\mathrm{span}\{\ket{0},\ket{1}\}$, we easily find that $\ket{\psi_0},\ket{\varphi_1},\ket{\psi_3},\ket{\varphi_3}\in\mathrm{span}\{\ket{0},\ket{1}\}$.
Projecting further onto $(0,0,1,\alpha)$ $(\alpha\in\mathbbm{C})$ and after careful analysis, one arrives at the conclusion that again one of the vectors spanning $V$ is of Schmidt rank 2.

Let us now come to the cases (ii) and (iii). As they are
analogous, for simplicity we consider only the
first of them. The corresponding conditions imply that
$\ket{\psi_0}\parallel\ket{1}$ and
$\ket{\psi_0},\ket{\varphi_3}\in\mathrm{span}\{\ket{0},\ket{1}\}$.
Then, it suffices to project onto $\ket{x}=(1,\alpha,0,0)$
$(\alpha\in\mathbbm{C})$ to see that
$\ket{\varphi_1}\in\mathrm{span}\{\ket{0},\ket{1}\}$, meaning that
we fall into the already discussed case when
$\ket{\psi_0},\ket{\varphi_1},\ket{\psi_3},\ket{\varphi_3}\in\mathrm{span}\{\ket{0},\ket{1}\}$
(see above).

Let us consider the case (iv) and project, for
instance, onto $\ket{x}=(1,0,\alpha,0)$ $(\alpha\in\mathbbm{C})$.
This, in particular, imposes the conditions that
$\ket{\psi_0'}\parallel\ket{\varphi_2'}$ or $\ket{\varphi_2'}=0$.
In the latter case,
$\ket{\varphi_2}\parallel\ket{\varphi_3}\parallel\ket{0}$ and one
sees immediately that $\ket{\Psi_3}$ is of Schmidt rank at most
2. Assume then that $\ket{\psi_0'}\parallel\ket{\varphi_2'}$ and
recall that either $\ket{\psi_0}$ and $\ket{\varphi_1}$ are
parallel or they both live in $\mathrm{span}\{\ket{0},\ket{1}\}$.
This implies that either $\ket{\varphi_1'}\parallel
\ket{\varphi_2'}$ or
$\ket{\varphi_1},\ket{\varphi_2}\in\mathrm{span}\{\ket{0},\ket{1}\}$.
In the first case,
$\ket{\Psi_3}\in\mathbbm{C}^4\ot\mathrm{span}\{\ket{0},\varphi_1'\}$,
while in the second one
$\ket{\Psi_3}\in\mathbbm{C}^4\ot\mathrm{span}\{\ket{0},\ket{1}\}$.
Hence, both cases are of rank at most 2. This completes
the proof. $\square$

Finally, let us follow Remark 1 and prove that any CES in $\mathcal{H}_n$
of dimension less than $n-1$ is a subspace of the
$(n-1)$-dimensional CES. For this purpose, we prove a bit more
general fact.\\

\noindent\textbf{Lemma 2.} {\it Any $r$-dimensional CES in
$\mathcal{H}_{m,n}$ with $r<(m+1)(n+1)$ is a subspace of an
$(r+1)$-dimensional CES.}\\

\noindent\textbf{Proof.} Let $V\subset\mathcal{H}_{m,n}$ denote an
$r$-dimensional CES and $M_V$ denoet a set constructed from linear
combinations of vectors from $V$ and all product vectors from
$\mathcal{H}_{m,n}$, that is,
\begin{equation}
M_V = \{\ket{\Psi} + \ket{e} \otimes \ket{f} \,\,|\,\,
\ket{\Psi} \in V, \ket{e} \in \mathbb{C}^n, \ket{f} \in
\mathbb{C}^m\}.
\end{equation}

We start by noting that $V$ can be extended to some
$(r+1)$-dimensional CES, denoted $V_2$, if
$M_V\subsetneq\mathcal{H}_{m,n}$. This is because if the set $M_V$ is
a proper subset of $\mathcal{H}_{m,n}$, we can take a vector
$\ket{\Psi}\in\mathcal{H}_{m,n}$ which does not belong to $M_V$
and define the subspace $V_2$ as a linear hull of that vector and
$V$. The construction of $M_V$ guarantees that $V_2$ is CES because
otherwise $\ket{\Psi}$ can be written as a linear combination of a
vector from $V$ and a product vector from $V_2$. Consequently,
$\ket{\Psi}$ has to be in $M_V$, contradicting the assumption that
$\ket{\Psi}\not\in M_V$.

Let us now show that for any CES of nonmaximal dimension [$\dim V
< (n-1)(m-1)$] the set $M_V$ is not the whole $\mathbb{C}^n
\otimes \mathbb{C}^m$. For this purpose, observe first that $M_V$
is a manifold. Every its element can be represented as a sum of an
element from $(n+m-1)$-dimensional manifold of separable vectors
from $\mathcal{H}_{m,n}$ and the $r$-dimensional CES $V$, and
hence the dimension of $M_V$ is less or equal to $r+n+m-1$. Therefore, for any $r<(m-1)(n-1)$, the dimension of $M_V$ is less than $mn$,
meaning that $M_V\subsetneq\mathcal{H}_{m,n}$. Consequently, $V$ can be extended to an $(r+1)$-dimensional CES. $\square$

By repeating the above procedure until
$r=(m-1)(n-1)$ we see that every CES of a nonmaximal dimension in
$\mathcal{H}_{m,n}$ is a subspace of some maximally dimensional
CES.

In conclusion, we see, via theorem 1 and both the above lemmas,
that all completely entangled subspaces of $\mathcal{H}_4$ of
dimension less than or equal to 3 have the property of spanning.
This, by virtue of the previous discussion, proves the following theorem. \\

\noindent{\bf Theorem 2.} {\it Let $W$ be a decomposable witness
acting on $\mathcal{H}_4$. Then, the following statements are
equivalent:}
\begin{description}
\item[(i)]{\it  $W=Q^{\Gamma}$, $Q\geq 0$, $r(Q)\leq 3$, and $\mathrm{supp}(Q)$ is
a CES supported in $\mathcal{H}_4$,
\item[(ii)] $P_W$ spans $\mathcal{H}_4$,
\item[(iii)] $W$ is optimal.}
\end{description}

Collecting together results obtained for $\mathcal{H}_2$ and
$\mathcal{H}_3$ in Ref. \cite{nasza} and the above one for
$\mathcal{H}_4$, it is tempting to conjecture that any CES $V$
such that $\mathrm{dim}V\leq n-1$ supported in $\mathcal{H}_n$ has the property of spanning. In other words, we conjecture that Theorem 2 holds for any DEW (\ref{form}) from $\mathbbm{M}_n$ as long as $r(Q)\leq n-1$.

Interestingly, it is easy to disprove this conjecture for systems
of unequal local dimensions. For this purpose, let us consider the
subspace of $\mathcal{H}_{m,n}$ with $n>m$, spanned by the
following vectors:
\begin{eqnarray}
\ket{\Psi_i}&=&\ket{0}\ket{i}-\ket{i}\ket{0}\qquad (i=1,\ldots,m-1)\nonumber\\
\ket{\Psi_i}&=&\ket{0}\ket{i}-\ket{1}\ket{i-1}\qquad (i=m,\ldots,n-2)\nonumber\\
\ket{\Psi_{n-1}}&=&\ket{0}\ket{n-1}-\ket{\psi_{\mathrm{ant}}},
\end{eqnarray}
where $\ket{\psi_{\mathrm{ant}}}$ is a state from the
antisymmetric subspace of $\mathbbm{C}^{m}\ot\mathbbm{C}^{m}$
orthogonal to all $\ket{\Psi_i}$ $(i=1,\ldots,m-1)$. Notice that
in the case $n=m+1$, one omits the second group of vectors.

Using the matrix representation of the above vectors and applying
analogous reasoning as in the Appendix, one finds that
$V=\spa\{\ket{\Psi_i}\}_{i=1}^{n-1}$ is a $(n-1)$-dimensional
subspace supported in $\mathcal{H}_{m,n}$ which does not contain
any product vectors. Then, there are two classes of product
vectors orthogonal to this subspace, that is,
\begin{equation}
(1,x_1,\ldots,x_{m-1})\ot(y_0,\ldots,y_{n-2},0)
\end{equation}
and
\begin{equation}
(0,x_1,\ldots,x_{m-1})\ot\ket{y},
\end{equation}
where now $\ket{y}$ can be arbitrary. Let us now take partial
conjugations of both classes. Clearly, the product vector
$\ket{0}\ket{n-1}$ is orthogonal to both of them and therefore the
above subspace does not have the property of spanning. To
illustrate the above construction with an example, let us consider
the case of $\mathcal{H}_{3,4}$. The above construction gives
$\mathcal{V}=\spa\{\ket{01}-\ket{10},\ket{02}-\ket{20},\ket{03}-(\ket{12}-\ket{21})\}$.
The product vectors orthogonal to $\mathcal{V}$ are given by
$(1,x_1,x_2)\ot(y_0,y_1,y_2,0)$
$(x_1,x_2,y_0,y_1,y_2\in\mathbbm{C})$ and $(0,x_1,x_2)\ot\ket{y}$
$(x_1,x_2\in\mathbbm{C},\ket{y}\in\mathbbm{C}^4)$. The vector
orthogonal to partial conjugations of both classes is
$\ket{0}\ket{3}$.

\section{Conclusion}
\label{Conclusion}

Let us here shortly summarize the obtained results and outline the
possibilities for further research.

Because entanglement witnesses are very useful tools in quantum
information theory, their characterization is of great interest.
Particularly important in this context is the notion of optimality
introduced in Ref. \cite{optimization}.

Very recently, some of us have studied decomposable witnesses
detecting entanglement in qubit-qunit systems and provided a
complete characterization of optimality in this case \cite{nasza}.
In the present paper we have treated several questions that arise
in other finite-dimensional Hilbert spaces. One of the most
interesting problems is whether optimal
decomposable witnesses exist such that the corresponding
$\mathcal{P}_W$s do not span the Hilbert space. Here we answer
this question positively by showing that for any Hilbert space
$\mathcal{H}_{m,n}$ with $m,n \geq 3$ there exist optimal DEWs
without the property of spanning.

Aiming at the generalization of the results of Ref.
\cite{nasza}, we have tried to distinguish all the CESs for which
the corresponding $\mathcal{P}_V$s do span $\mathcal{H}_{m,n}$.
This tells us for which DEWs of form (\ref{form}) the
implications (i) and (ii) (see Sec. \ref{Preliminaries}) become
equivalences. We have proven that under a certain
condition, which is conjectured to be always satisfied,  all
$(n-1)$-dimensional CESs supported in $\mathcal{H}_{n}$ have the
property of spanning. This result obviously extends to any
CES being a proper subspace of such an $(n-1)$-dimensional CES
supported in two-qunit Hilbert space. We have
proven that any $r$-dimensional CES can be extended to an
$(r+1)$-dimensional one, meaning that all CESs in
$\mathcal{H}_{n}$ of a dimension less than $n-1$
are subspaces of $(n-1)$-dimensional CESs. We have applied this statements to the two-ququart case and shown
that any CES of dimension $\mathrm{dim}V\leq 3$ supported in
$\mathcal{H}_4$ has the property of spanning. This, together with
the results obtained in Ref. \cite{nasza}, allows us to conjecture
that any CES of dimension less than or equal to $n-1$ and supported in
$\mathcal{H}_n$ has this property. Certainly this conjecture
cannot hold if the local dimensions are different, as we provide
examples of $(n-1)$-dimensional subspaces without the property of
spanning supported in $\mathcal{H}_{m,n}$ $(3\leq m < n)$.

Clearly, still much has to be done to complete the
characterization of witnesses. In particular, it remains to prove the above conjecture and determine whether all CESs of
$\mathcal{H}_{m,n}$ $(m,n\geq 3)$ have the property of spanning.
This would allow to find all the instances when (i) and (ii) (see
Sec. \ref{Preliminaries}) are equivalences. The more challenging
task, however, would be to ask a similar question in the case of
indecomposable witnesses.

\acknowledgments{Discussions with Seung-Hyeok Kye, Pawe\l{}
Mazurek, \L{}ukasz Skowronek, and Julia Stasi\'nska are gratefully
acknowledged. This work was supported by EU IP AQUTE, Spanish
MINCIN Project FIS2008-00784 (TOQATA), Consolider Ingenio 2010
QOIT, EU STREP Project NAMEQUAM, and the Alexander von Humboldt
Foundation. M. L. acknowledges NFS Grant No. PHY005-51164.}

\appendix

\section{Proving that considered subspaces are CES}

Let us first recall the vector-matrix isomorphism. Precisely, to
any vector $\mathcal{H}_{m,n}\ni
\ket{\psi}=\sum_{ij}a_{ij}\ket{ij}$ there correspond
a complex $m\times n$ matrix
$A=\sum_{ij}a_{ij}\ket{i}\!\bra{j}$ ({\it and vice versa}).

\subsection{The symmetric case}

We consider an $m$-dimensional subspace $V$ in $\mathbb{C}^m
\otimes \mathbb{C}^m$ spanned by vectors:
\begin{displaymath}
 \ket{\Psi_i} = \ket{0} \ket{i} - \ket{i}\ket{0} \qquad (i = 1, \dots, m-1),
 \label{baza_mm}
\end{displaymath}
and a vector $\ket{\Psi_m} \in \mathrm{span}(\ket{1}, \dots,
\ket{m-1})^{\otimes 2}$.

We prove that the only product vector in this subspace is zero.
For this purpose, let us take a linear combination of all vectors
spanning $V$, which, in terms of the above vector-matrix
correspondence, reads
\begin{equation}\label{A1}
 \left[ \begin{array}{c|ccc}
 0 & \alpha_1 & \dots & \alpha_{m-1} \\
 \hline
 -\alpha_1 & & & \\
 \vdots & & \alpha_m \widetilde{A}_m & \\
 -\alpha_{m-1} &  &  &
 \end{array} \right]
 \end{equation}
where $r(A_m) > 1$. There exists a product vector in $V$ iff there
exist nonzero $\alpha$s such that the above matrix is of rank 1.

First observe that $\ket{\Psi_m}$ affects neither the first column
nor the first row of (\ref{A1}). The principal minors of the
matrix (\ref{A1}) formed by the sets of indices $\{0,i\}$, $i \in
\{1,m-1\}$ are then equal $\alpha_1^2, \dots, \alpha_{m-1}^2$. If
the matrix is to be of rank 1, all of them have to be zero, which
implies that $\alpha_1 = \dots = \alpha_{m-1} = 0$. The last
coefficient $\alpha_m$ has also to be zero, because $r(A_m )>1$.
This finishes the proof.

\subsection{The general case}

Now we consider an $n$-dimensional subspace in $\mathbb{C}^m
\otimes \mathbb{C}^n$ spanned by vectors $\ket{\Psi_i}$ $(i=1,
\dots, n-1)$ given by Eqs. (\ref{PsiI}) and (\ref{PsiII}), but in
this case the vector $\ket{\Psi_{m}}$ is fixed and defined by
(\ref{macierzB}). The combination matrix has the form
\begin{equation}
 \left[ \begin{array}{c|ccc|cccc}
 0 & \alpha_1 & \dots & \alpha_{m-1} & \alpha_{m+1} & \alpha_{m+2} & \dots & \alpha_n \\
 \hline
 -\alpha_1 & & & x & -\alpha_{m+2} & \dots & \alpha_n & 0 \\
 \vdots & & \alpha_m A_m & & 0 & \dots & \dots & 0 \\
 -\alpha_{m-1} &  &  & & 0 & \dots & \dots & 0
 \end{array} \right],
 \end{equation}
where $x=\alpha_m a_{0,m-2}-\alpha_{m+1}$.

Again, by calculating the principal minors of the combination matrix
formed by the sets of indices $\{0,i\}$, $i \in \{1,m-1\}$, one
gets $\alpha_0 = \dots = \alpha_{m-2} = 0$. The block formed by
the last $m-1$ rows and the columns of indices $1, \dots m-1$ is
affected by $\ket{\Psi_m}$ and $\ket{\Psi_{m+1}}$, but always has
rank $m-1$ if only $\alpha_m \ne 0$, so $\alpha_m = 0$. Now only the two first rows are nonzero in the combination matrix. We consider minors of size 2 formed by these rows and columns of indices $i-1$, $i$. Starting with $i=m+1$, one gets
$\alpha_{m+1}^2=0$. Taking $i=m+2$, one gets $\alpha_{m+2}^2=0$. In
this manner, one can show that $\alpha_{m+1} = \dots = \alpha_n =
0$, completing the proof.

\end{document}